\documentstyle[aps,prb,eqsecnum,amsfonts,epsf,multicol]{revtex}

\begin{document}

\newcommand{\uprule}{\end{multicols}
\noindent \vrule width3.375in height.2pt depth.2pt 
\vrule height.5em depth.2pt \hfill \widetext }
\newcommand{\downrule}{\indent \hfill \vrule depth.5em height0pt 
\vrule width3.375in height.2pt depth.2pt 
\begin{multicols}{2} \narrowtext}

\draft

\title{Hubbard model with $SU\left( 4\right) $ symmetry}
\author{Eugene Pivovarov}
\address{California Institute of Technology, Pasadena, California 91125}
\date{\today }
\maketitle

%%%%%%%%%%%%%%%%%%%%%%%%%%%%%%%%%%%%%%%%%%%%%%%%%%
\begin{abstract}
In the model considered, the nonlocal interaction of the fermions in 
different sublattices of a bipartite lattice is introduced. It can also 
be regarded as local interaction of fermions with opposite ``hypercharge''. 
The corresponding term in the Hamiltonian is $SU\left( 4\right) $-%
invariant and appears to be the most tractable version of the $SO\left( 5
\right) $-invariant model that unifies antiferromagnetic and $d$-wave 
superconducting order parameters. The model has been studied primarily in 
the weak interaction limit and in the mean field approximation. Near the 
half-filling the antiferromagnetic critical temperature has a peak. 
However, the superconducting transition takes place when the Fermi surface 
crosses the area where the density of states is of order of inverse coupling 
coefficient. Thus, in mean-field approximation, there exist an interval of 
values of the chemical potential, for which the system is a superconductor 
for arbitrary high temperatures. The temperature dependence of specific heat, 
Hall coefficient, and DC conductivity in the normal phase agrees with that 
experimentally observed in high-$T_c$ cuprates.
\end{abstract}


\begin{multicols}{2}

%%%%%%%%%%%%%%%%%%%%%%%%%%%%%%%%%%%%%%%%%%%%%%%%%%
\section{Introduction}

Among the properties of the high-$T_c$ cuprates that have long defied
explanation are the proximity of the antiferromagnetic (AF) and $d$-wave
superconducting (dSC) phases below the critical temperature and the
anomalous temperature dependence of kinetic and thermodynamic quantities
above the transition point. A recently proposed concept of an $SO\left(
5\right) $ symmetry between AF and dSC phases \cite{Zhang} aims to explain
the former as well as the resonance mode observed in spin-flip neutron
scattering on YBCO.\cite{Fong} Several groups \cite{Scalapino,Henley,Burgess} 
have constructed microscopic models with exact $%
SO\left( 5\right) $ symmetry, and it has been argued \cite{Meixner} that
the two-dimensional (2D) Hubbard model has approximate $SO\left( 5\right) $
symmetry.

The properties of the microscopic models with symmetries higher than 
$SO\left( 5\right) $ were also 
investigated earlier. In Refs.\ \onlinecite{Affleck85,Affleck} the large-$n$ 
limit of 
$SU\left( n\right) $ model has been studied and $1/n$ expansion has been 
applied. It has been found that in the strong coupling limit the ground 
state breaks translational symmetry and represents a density wave in which 
each site forms a dimer with one of its nearest neighbors. As the doping 
increases, a ``kite'' state with charge-density wave and no charge gaps 
forms. In the weak interaction limit the ``flux'' state with full 
translational symmetry and gap vanishing at discrete points in momentum 
space was predicted. However, it was shown that at large $n$ the ground 
state does not have off-diagonal long range order.

In Ref.\ \onlinecite{Markiewicz} an $SO\left( 6\right) $ model has been 
suggested, in 
which AF, dSC, and flux phase are unified. This and the subsequent work 
\cite{Markiewicz2} have shown that the pinning of the Fermi level near a Van 
Hove singularity can explain the observed stripe phases \cite{Tranquada} in 
cuprate superconductors.

This paper presents a study of a relatively simple two-leg ladder model that 
possesses $SU\left( 4\right) $ symmetry, which is higher than 
$SO\left( 5\right) $ symmetry. The introduced interaction favors the state, 
in which each rung is occupied by the pairs of fermions corresponding to the
opposite legs of the ladder, or ``flavors''. This model is associated 
directly to a 2D bipartite lattice, since each rung can be related to a site 
in one of the sublattices. Then one leg will be simply this sublattice, while 
the other leg will correspond to the second sublattice (Fig.\ \ref{Fig1}). 
The choice of such a correspondence 
is not unique and it affects the form that the interaction takes after the
transition from the two-leg ladder model to an equivalent 2D bipartite lattice.

%%%%%%%%%%%%%%%%%%%%%%%%%%%%%%%%%%%%%%%%%%%%%%%%%%
\begin{figure}
\epsfxsize =3.375 in
\epsffile{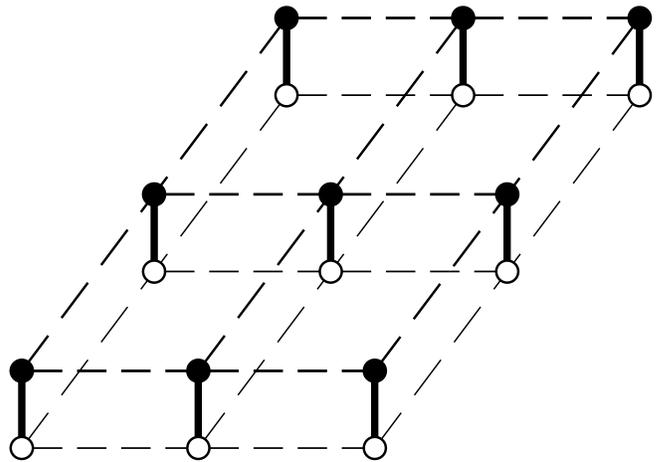}
\narrowtext
\caption{2D two-leg ladder: black and white circles comprise the first and 
the second legs, respectively. Solid lines denote rungs.}
\label{Fig1}
\end{figure}
%%%%%%%%%%%%%%%%%%%%%%%%%%%%%%%%%%%%%%%%%%%%%%%%%%

The kinetic term on the two-leg ladder includes next-neighbor 
hopping along the legs. For the equivalent 2D bipartite lattice, in general 
case, the hopping between remote sites will appear on the second sublattice. 
However, there exist \cite{Henley} such a correspondence between the ladder and 
the bipartite lattice so that the kinetic term on the bipartite lattice will 
include only next-neighbor hopping. Furthermore, the 
kinetic term will be $SU\left( 4\right) $-invariant only if each of the 
sublattices is bipartite as well. Then the corresponding two-leg ladder will 
be bipartite too, and its rungs can be enumerated so 
that even and odd rungs will belong to different subladders. It is the 
\textit{bipartite 2D two-leg ladder} that is considered in this paper.

The studied model is of special interest not only because of its probable  
relationship 
to high-$T_c$ cuprates, but also due to the possibility to synthesize such 
ladders and explore them experimentally.\cite{Dagotto} Therefore, one of the 
goals of the presented work is to  
derive the analytical expressions for the critical temperature of the 
transition, as well as for specific heat and for components of DC conductivity 
tensor in normal state. The calculation is performed in weak coupling limit 
and mean-field approximation, however, some of the results can be 
qualitatively extended to the strong coupling regime.

In this regime there are six possible phases that correspond to 15 generators 
of the algebra and the Casimir operator. The values of the operators 
alternate their signs at the nearest neighbor rungs, although not all of the 
phases are density waves. In presence of symmetry breaking terms, only AF and 
dSC states remain.

A peculiarity of the constructed model is that the interaction part of the
Hamiltonian is being split into two terms with $SO\left( 4\right) \times U
\left( 1\right) $ symmetry that
renormalize independently. Consequently, these terms correspond both to
different conditions on phase transition and to the different kind of the
transition (AF or dSC). In particular, it will be shown that in mean-field 
approximation the possibility of the dSC transition is determined by the value 
of chemical potential, but not by temperature, and therefore, \textit{the
system can be in the dSC state at arbitrary high temperatures}. This property 
is the direct consequence of the presence of $SU\left( 4\right) $-invariant 
interaction in the model that attracts the fermions of different 
``flavors''.

%%%%%%%%%%%%%%%%%%%%%%%%%%%%%%%%%%%%%%%%%%%%%%%%%%
\section{Model}

In a two-leg ladder model each rung can have up to four fermions,
differing in spin and ``flavor''. This model is related to an equivalent 2D
bipartite lattice if we assume that one flavor, $c$, corresponds to fermions
on the first sublattice, while the other flavor, $d$, is a linear
combination of the fermions on the second sublattice, $d_\sigma \left( 
{\mathbf r}_j\right) =\sum_{{\mathbf r}^{\prime }}\phi \left( {\mathbf r}_j-%
{\mathbf r}^{\prime }\right) c_\sigma \left( {\mathbf r}^{\prime }\right) $,
where $\phi \left( {\mathbf r}\right) =0$ for $x+y$ even and $j$ is labeling
the sites within the first sublattice. In the important case \cite{Henley} 
when $\phi \left( {\mathbf r}\right) =\left( 4/\pi \right) \left(
x^2-y^2\right) ^{-1}$ for $x+y$ odd, the operator $c_{{\mathbf k}%
\sigma }d_{{\mathbf k}\sigma }$ takes the form of the dSC order parameter $%
\Delta _{\mathbf k}=g_{{\mathbf k}}c_{{\mathbf k}\sigma }c_{-{\mathbf k}%
\,\sigma }$, where $g_{{\mathbf k}}=\,\text{sign}\left( \cos k_x-\cos
k_y\right) $ and the annihilation operators $c_{{\mathbf k}\sigma }$ and 
$c_{-{\mathbf k}\,\sigma }$ act on different sublattices. 

The definition above leads to a number of important consequences. 
First, we can regard $d$-particles as well-defined fermions, since their 
creation and annihilation operators obey canonical commutation relations. 
Second, the total number of $d$-fermions is equal to the number of fermions 
on the second sublattice. Third, in 2D case the contribution from the second 
sublattice to kinetic energy takes the same form be it written in 
terms of initial $c$-operators or newly defined $d$-operators. Thus, we 
conclude that $d$-fermions are the alternative representation of the second 
sublattice of the 2D bipartite lattice.

As it was mentioned in the introduction, we will assume that the two-leg 
ladder is bipartite. 
In order to simplify the construction of the Hamiltonian with explicit $%
SU\left( 4\right) $ symmetry, let us group the fermion operators on a rung
into a 4-component operator 
\begin{equation}
\Psi _j^{\dagger }=\left( 
\begin{array}{llll}
c_{j\uparrow }^{\dagger }, & c_{j\downarrow }^{\dagger }, & \left( -1\right)
^jd_{j\uparrow }, & \left( -1\right) ^jd_{j\downarrow }
\end{array}
\right) .
\label{Psi}
\end{equation}

The terms that include only the scalar products of such operators
are $SU\left( 4\right) $-invariant. Those that involve 
the antisymmetric inner product $\Psi _{j\alpha
}^{\dagger }E_{\alpha \beta }\Psi _{j\beta }^{\dagger }$ reduce the symmetry
of the group to $Sp\left( 4\right) $, or equivalently, $%
SO\left( 5\right) $. As we will see below, such terms bring about leg-to-leg 
hopping within a rung.

Consider a model with the Hamiltonian 
\begin{equation}
{\mathcal H}={\mathcal H}_{\text{kin}}+{\mathcal H}_{\text{int}}+{\mathcal H}_{%
C}+{\mathcal H}_{\text{chem}}  \label{model0}
\end{equation}
Here the kinetic (hopping) and the scalar interaction terms are $SU\left(
4\right) $-invariant: 
\begin{eqnarray}
{\mathcal H}_{\text{kin}} &=&-t\sum_{\left\langle i,j\right\rangle \sigma
}\left( c_{i\sigma }^{\dagger }c_{j\sigma }+d_{i\sigma }^{\dagger
}d_{j\sigma }\right) =-t\sum_{\left\langle i,j\right\rangle }\Psi
_i^{\dagger }\Psi _j\\
{\mathcal H}_{\text{int}} &=&U\sum_jY_j^2
=U\sum_j\left( \Psi _j^{\dagger }\Psi _j-2\right) ^2,  \label{interaction}
\end{eqnarray}
where the ``hypercharge'' operator $Y_j=n_j^{(c)}-n_j^{(d)}%
=\Psi _j^{\dagger }\Psi _j-2$, $n_j^{(c)}=\sum_%
\sigma c_{j\sigma }^{\dagger }c_{j\sigma }$, and $n_j^{(d)}=\sum_\sigma
d_{j\sigma }^{\dagger }d_{j\sigma }$. Thus, different flavors have opposite 
hypercharge. Also note that the kinetic term is invariant only globally, 
since it contains scalar products of the operators on different rungs, 
while the interaction term is locally invariant as well.

The $SU\left( 4\right) \rightarrow SO\left( 4\right) \times U\left( 1\right) 
$ breaking terms are chemical potential and Coulomb interaction. The latter
can be regarded also as dSC~--\ AF anisotropy, as it can be expressed in
terms of the square of the local spin operator:\cite{Fradkin}
\begin{eqnarray}
{\mathcal H}_{\text{chem}} &=&-\mu _0\sum_jn_j,  \label{mu_h} \\
{\mathcal H}_{C} &=&-\frac{4g}3\sum_j\left( \left| {\mathbf S}%
_j^{(c)}\right| ^2+\left| {\mathbf S}_j^{(d)}\right| ^2+n_j\right)   \nonumber
\\
&=&g\sum_j\left[ \left( n_j^{(c)}-1\right) ^2+\left( n_j^{(d)}-1\right)
^2-2\right] .  \label{Coulomb}
\end{eqnarray}
Here $n_j=n_j^{(c)}+n_j^{(d)}$. One can also introduce the $SU\left(
4\right) \rightarrow SO\left( 5\right) $ symmetry breaking terms so that
their combination with Eqs.~(\ref{mu_h},\ref{Coulomb}) will finally reduce
the symmetry to $SO\left( 3\right) \times U\left( 1\right) $. Such terms are
the $SO\left( 5\right) $ ``rotor'', or Casimir, operator and the inter-rung
leg-to-leg hopping:

\uprule

\begin{equation}
{\mathcal H}_{\text{rot}} =\frac 1{2\chi }\sum_j\sum_{a<b}{\mathcal L}%
_{jab}^2 
=\sum_j\left[ \frac 1{2\chi }\left( n_j^{(c)}-1\right) \left(
n_j^{(d)}-1\right) +\frac 2\chi {\mathbf S}_j^{(c)}\cdot {\mathbf S}%
_j^{(d)}\right] ,
\end{equation}

\downrule

\begin{eqnarray}
{\mathcal H}_{\text{hop}} &=&-t^{\prime }\sum_{j\sigma }\left[ 1+\kappa
\left( n_{j,-\sigma }^{(c)}-n_{j,-\sigma }^{(d)}\right) ^2\right] c_{j\sigma
}^{\dagger }d_{j\sigma } \nonumber \\
&&+h.\ c. \nonumber \\
&=& -t^{\prime }\sum_{j\sigma }\left( 1+\kappa \Psi _j^{\dagger }\Psi
_j\right) \Psi _j^{\dagger }E\Psi _j^{\dagger }+h.\ c.,  \label{hop}
\end{eqnarray} 

\noindent 
where ${\mathcal L}_{jab}$ are the generators of $SO\left( 5\right) $
symmetry defined below in Eq.~(\ref{L}), ${\mathbf S}^{(c)}=\frac 12c^%
{\dagger }\bbox{\sigma }c$, and ${\mathbf S}^{(d)}=\frac 12d^{\dagger }%
\bbox{\sigma }d$. In the absence of leg-to-leg
hopping Eq.~(\ref{hop}), the total hypercharge of the system 
$\sum_jY_j$ is a conserved quantity.

Finally, there exist a class of terms that break symmetry to $SU\left( 3
\right) $. They correspond to the presence of certain inhomogeneous magnetic
field that takes different values on the sublattices. An example of such a
term is $-{\mathbf H}^{(c)}\cdot {\mathbf S}^{(c)}$.

The free energy spectrum of both $c$ and $d$-fermions is $\varepsilon \left(
{\mathbf k}\right) =-2t\left( \cos k_x+\cos k_y\right) $. The interaction
term in Eq.~(\ref{model0}) can be written as a sum of 15 generators of $SO
\left( 6\right) \cong SU\left( 4\right) $ so that ${\mathcal H}_{\text{int}}$ 
becomes $\sum_j\hat{\mathcal H}_{\text{int},j}$ with 
\begin{eqnarray}
\hat{\mathcal H}_{\text{int},j} &=&U\left[ 4-\frac
15\sum_{a=0}^5\sum_{b=a+1}^5\left( \Psi _j^{\dagger }M_{ab}\Psi _j\right)
^2\right]   \nonumber \\
&=&4U \left( 1-\frac 15\sum_{a=1}^5{\mathcal N}_{ja}^2-\frac
15\sum_{a=1}^5\sum_{b=a+1}^5{\mathcal L}_{jab}^2\right) .  \label{expansion}
\end{eqnarray}
Here $M_{ab}$ are the generators of the matrix representation of $SO\left(
6\right) $ that acts on the space of 4-by-4 matrices by conjugation, $%
{\mathcal L}_{ab}$ are the generators of the representation of $SO\left( 5%
\right) $ and ${\mathcal N}_a$ is the corresponding superspin:
\begin{mathletters} 
\begin{eqnarray}
{\mathcal L}_{ab} &=&\frac 12\Psi ^{\dagger }M_{ab}\Psi \quad \text{for }%
a,b=1\,...\,5,  \label{L} \\
{\mathcal N}_a &=&\frac 12\Psi ^{\dagger }M_{0a}\Psi \quad \text{for }%
a=1\,...\,5.
\end{eqnarray}
\end{mathletters}

It is convenient to choose the following representation for $M_{ab}$:
\cite{Scalapino}
\begin{eqnarray*}
M_{0a} &=&\left( 
\begin{array}{cc}
\sigma _a & 0 \\ 
0 & \sigma _a^T
\end{array}
\right) ,\ a=1,2,3, \\
M_{04} &=&\left( 
\begin{array}{cc}
0 & -i\sigma _y \\ 
i\sigma _y & 0
\end{array}
\right) ,\ M_{05}=\left( 
\begin{array}{cc}
0 & \sigma _y \\ 
\sigma _y & 0
\end{array}
\right) , \\
M_{ab} &=&-\frac i2\left[ M_{0a},M_{0b}\right] ,\ a,b=1\,...\,5.
\end{eqnarray*}

The physical meaning of the components is ${\mathcal N}_a=\left(
m_x,m_y,m_z,\text{Re}\Delta _{{\mathbf Q}},\text{Im}\Delta _{{\mathbf Q}%
}\right) ^T$, where antiferromagnetic order parameter ${\mathbf m} 
=\frac 12\left( c^{\dagger }\bbox{\sigma }c-d^{\dagger }\bbox{\sigma }%
d\right) $, $\Delta _{\mathbf Q} =\Delta \exp \left( %
-i{\mathbf Q}\cdot {\mathbf r}\right) $, dSC order parameter 
$\Delta =ic\sigma _yd$, ${\mathbf Q}=\left( \pi ,\pi \right) $, and $%
{\mathcal L}_{ab}$ incorporates rung spin operator ${\mathbf S} 
={\mathbf S}^{(c)}+{\mathbf S}^{(d)}$, $\pi $--operator $\bbox{\pi }^%
{\dagger } =-\frac 12c^{\dagger }\bbox{\sigma }\sigma _yd^{\dagger }$, and 
electric charge density $Q=\frac 12\left( n^{(c)}+n^{(d)}\right) -1$. In the
absence of $SU\left( 4\right) $ symmetry
breaking, the components of ${\mathcal L}_{ab}$ can also evolve into order
parameters, such as ferromagnetic order parameter and $\pi $-wave
superconducting order parameter. Note that if we formally replace $d$-
operators in $\Delta _{\mathbf Q}$ by $c$-operators, the result will 
coincide with the expression for the $\eta$-operator that generates $SO(3)$ 
pseudospin symmetry in the standard Hubbard model.\cite{Yang}

All 15 generators of $SU\left( 4\right) $ algebra can be regarded as components 
of a \textit{superspin} that transforms according to the adjoint representation 
of $SU\left( 4\right) $. Although the superspin is not the order parameter, it 
is directly related to one.

It is known from the theory of the 2D standard Hubbard model \cite{Fradkin} %
that at half-filling below the transition point the density-wave state has 
lower ground
energy than the spatially homogeneous state if the lattice is bipartite. 
In such a state the components of the superspin 
will alternate the sign at even and odd rungs. Thus, if the ladder (or 
each of the sublattices of the equivalent bipartite lattice) is bipartite,
charge-density wave (CDW), spin-density wave (SDW), and dSC state are the
only actually possible ordered phases at half-filling. In the \textit{pure} 
$SU\left( 4\right) $ theory\cite{note}
there are totally six phases that can be classified by the type of
order and the parity with respect to the exchange of $c$ and $d$-particles.
The table below displays the components of the superspin that vary as 
$\cos \left( {\mathbf Q}\cdot {\mathbf r}\right) $ for CDW and SDW states and 
the actual order parameters for the dSC states, according to such a 
classification: 
\begin{equation}
\begin{tabular}{|c|c|c|c|}
\hline
O$\text{rder}$ & CDW & SDW & dSC \\ \hline
$\text{odd parity}$ & $n^{(c)}-n^{(d)}$ & ${\mathbf S}^{(c)}-{\mathbf S}^{(d)}%
$ & $\Delta ,\Delta ^{\dagger }$ \\ 
$\text{even parity}$ & $n^{(c)}+n^{(d)}-2$ & ${\mathbf S}^{(c)}+{\mathbf S}%
^{(d)}$ & $\bbox{\pi } ,\bbox{\pi }^{\dagger }$ \\ \hline
\end{tabular}
\end{equation}
Note that the odd parity CDW phase takes place only when the coupling $U$ is 
negative, while the rest only when it is positive.

In the dSC states the components of the superspin vary as 
$\cos \left( {\mathbf Q}\cdot {\mathbf r}\right) $, but due to the presence 
of $\left( -1\right) ^j$ factor in Eq.~(\ref{Psi}), these components (such as 
$\Delta _{\mathbf Q}$) become naturally related to the quantities that are 
constant everywhere in the dSC state (such as $\Delta$). The latter are the 
order parameters. In the case of CDW and SDW states, the order parameter is 
the amplitude of the variation, or the difference between the value of the 
superspin on even and odd rungs.

There are also 16 eigenstates of $\hat{\mathcal H}_{\text{int}}$ [Eq.~(\ref{%
expansion})] that can be labeled by the eigenvalues of ${\mathcal N}^2%
=\sum_a {\mathcal N}_a^2$, ${\mathcal L}^2=\sum_a\sum_{b>a}%
{\mathcal L}_{ab}^2$, rung spin component $S_z$, charge density $Q$, and 
hypercharge $Y$ (Table~\ref{table2}). The ground state of $%
\hat{\mathcal H}_{\text{int}}$ is 6-fold degenerate, consisting of the state 
$\left| \Omega %
\right\rangle =\frac 1{\sqrt{2}}\left( c_{\uparrow }^{\dagger }d%
_{\downarrow }^{\dagger}-c_{\downarrow }^{\dagger }d_{\uparrow }^%
{\dagger }\right) \left|0\right\rangle $ and its five transformations by 
the components of the $SO\left( 5\right) $ superspin. Therefore, 
in the ground state of the total interaction term $%
{\mathcal H}_{\text{int}}$, each rung is occupied only by $c-d$ pairs.

The degeneracy of the ground state of ${\mathcal H}_{\text{int}}$ at 
half-filling includes the contributions from the states, in which every rung 
of the two-leg ladder is occupied by one $c-d$ pair, and the states with 
some of the rungs being fully occupied and some empty, or in other words, 
with four-particle $c-d$ Cooper pairs. In the absence of $c-d$ Cooper pairs, 
the degeneracy of the ground state is one of standard Hubbard Hamiltonian, 
$2^N$, where $N$ is a number of fermions on the lattice. Taking into account 
that $k$ rungs can be empty and $k$ rungs full, we can find the total 
degeneracy and express it in terms of hypergeometric function 
$F\equiv \,_2F_1\left( a,b;c;x\right) $:
\begin{eqnarray}
\sum_{k=0}^{N/4}2^{N-4k}\left( 
\begin{array}{c}
N/2 \\ 
2k
\end{array}
\right) \left( 
\begin{array}{c}
2k \\ 
k
\end{array}
\right) &=& 2^NF\left( \frac 12-\frac N4,-\frac N4;1;\frac 14
\right) , \nonumber 
\end{eqnarray}
and for large values of $N$ this expression can be roughly approximated as 
$5^{N/8}\left( 8/N\right) ^{1/2}$.

In the strong coupling limit, the ground state is approximately one of $%
{\mathcal H}_{\text{int}}$ and we can derive the analog of the $t-J$ model
by computing the second-order correction to the kinetic term (as zero and
first orders vanish in the ground state). Using the identity ${\mathbf M}%
_{\alpha \beta }\cdot {\mathbf M}_{\gamma \delta }=4\delta _{\alpha \delta
}\delta _{\beta \gamma }-\delta _{\alpha \beta }\delta _{\gamma \delta }$
and taking into account that $Y_j=0$ in the ground state of $%
{\mathcal H}_{\text{int}}$, we find: 
\begin{eqnarray}
{\mathcal H}_{t-J} &=&{\mathcal H}_{\text{kin}}+J\sum_{\left\langle
i,j\right\rangle }\left( {\mathcal M}_i\cdot {\mathcal M}_j-{\mathbf t}%
_i^{\dagger }\cdot {\mathbf t}_j\right) ,
\label{t-J}
\end{eqnarray}
where $J=t^2/U$, ${\mathcal M}_{j,ab}=\frac 12\Psi _j^{\dagger }M_{ab}\Psi _j$%
, and ${\mathbf t}_{j,ab}=\frac 12\Psi _jM_{ab}\Psi _j$. Note that in the given 
representation for $M_{ab}$ some of the components of ${\mathbf t}_j$ 
vanish. The third term in Eq.~(\ref{t-J}) has a physical interpretation as 
pair hopping.

%%%%%%%%%%%%%%%%%%%%%%%%%%%%%%%%%%%%%%%%%%%%%%%%%%
\section{Critical temperature}

In this section we will use the temperature Green functions technique to derive the
expression for the critical temperature of the phase transitions in the 
system with the Hamiltonian Eq.~(\ref{model0}). The bare fermion
Green function is ${\mathfrak G}_{\alpha \beta }^{\left( 0\right) }\left( %
{\mathbf k},\omega \right) =\delta _{\alpha \beta }/\left[ i\omega
_n-Y_\alpha \left( \varepsilon _{{\mathbf k}}-\mu \right) \right] ,$ $\omega
_n=2\pi \left( n+1/2\right) k_BT,$ where Greek indices run from 1
to 4 and correspond to $c_{\uparrow }$, $c_{\downarrow }$, $d_{\uparrow
}^{\dagger }$, and $d_{\downarrow }^{\dagger }$, respectively; $Y_\alpha =+1$
for $\alpha =1,2,$ and $-1$ for $\alpha =3,4$. First, consider the case when 
$g=0$. The antisymmetrized bare interaction vertex corresponds to the factor 
$-{\mathfrak T}^{\left( \gamma _1,\gamma _2;\gamma _3,\gamma _4\right) }\left(
k_1,k_2;k_3,k_4\right) =U\left( \delta _{\gamma _1\gamma
_4}\delta _{\gamma _2\gamma _3}-\delta _{\gamma _1\gamma _3}\delta _{\gamma
_2\gamma _4}\right) $, where the notation $k=\left( {\mathbf k},\omega
\right) $ has been used.

%%%%%%%%%%%%%%%%%%%%%%%%%%%%%%%%%%%%%%%%%%%%%%%%%%
\begin{figure}
\hspace{0.125 in}
\epsffile{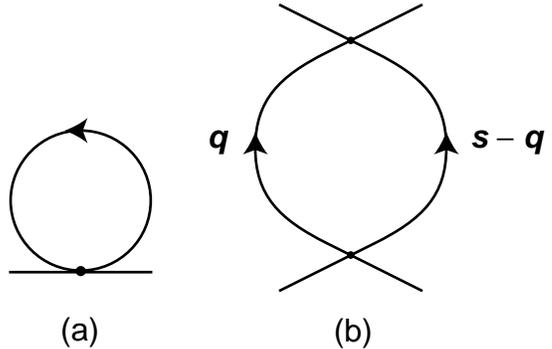}
\narrowtext
\caption{(a) Self-energy in one-loop order. 
(b) Correction to interaction vertex.}
\label{Fig2}
\end{figure}
%%%%%%%%%%%%%%%%%%%%%%%%%%%%%%%%%%%%%%%%%%%%%%%%%%

The chemical potential is $\mu =\mu _0+Y_\alpha \Sigma ^{\left(
\alpha ,\alpha \right) }\left( 0\right) ,$ $\left( \alpha =1,...,4\right) $
where $\Sigma ^{\left( \alpha ,\beta \right) }\left( {\mathbf k},\omega
_n\right) $ is the self-energy. The latter can be computed to one-loop order
[Fig.\ \ref{Fig2}(a)] and turns out to be a constant, $\Sigma ^{\left( \alpha ,
\beta \right)
}\left( {\mathbf k},\omega _n\right) =\Sigma ^{\left( \alpha ,\beta \right)
}\left( 0\right) $. Therefore, in one-loop order the Green function does not
change, since the correction to the single particle energy and one to the 
chemical potential cancel, hence, the Fermi 
level remains $\varepsilon _F=\mu -Y_\alpha \Sigma ^{\left( \alpha ,\alpha %
\right) }\left(0\right) =\mu _0$. The corresponding density of states can be 
found from the expression for the free-particle energy $\varepsilon 
_{\mathbf k}=-2t\left( \cos k_x+\cos k_y\right) $ and near the zero energy 
level $\varepsilon _{\mathbf k}=0$ it diverges logarithmically: 
\begin{equation}
\rho \left( \varepsilon \right) \simeq \text{sign}\left( \varepsilon \right) %
\frac%
1{\left( 2\pi \right) ^2}\frac 1{2t}\left[ 4\ln \left( \frac{\left| \varepsilon
\right| }{8t}\right) -2.75\right] ,\ \left| \varepsilon \right| \ll 2t.
\label{dos}
\end{equation}

Next we allow $g$ to take a small nonzero value. Then it is necessary to 
split ${\mathcal H}_{\text{int}}$
Eq.~(\ref{interaction}) into a sum of two terms so that one of them will be 
similar to the Coulomb term Eq.~(\ref{Coulomb}). Introduce two vectors 
$P^{(1)}$ and $P^{(2)}$ defined as follows: $P_\alpha ^{(1)}=1$ for $\alpha %
=1,2$ and $0$ for $\alpha =3,4$ and $P_\alpha ^{(2)}=1-P_\alpha ^{(1)}$.
Then each term in Eq.~(\ref{expansion}) can be represented as $\left( \Psi _j
^{\dagger }M_{ab}\Psi _j\right) ^2=\tilde{h}_{j,ab}^{\left( 1\right)
}+\tilde{h}_{j,ab}^{\left( 2\right) },$ where $\tilde{h}_{j,ab}^{\left( 1\right)
}={\mathcal M}_{j,ab}^{(1)2}+{\mathcal M}_{j,ab}^{(2)2}$, $\tilde{h}_{j,ab}^{%
\left( 2\right)}=2{\mathcal M}_{j,ab}^{(1)}{\mathcal M}_{j,ab}^{(2)}$, 
${\mathcal M}_{j,ab}^{(i)}=\Psi _j^{\dagger}M_{ab}^{(i)}\Psi _j$, 
and $M_{ab}^{(i)}\equiv M_{ab}\cdot P^{(i)}$. Thus,
\begin{eqnarray}
{\mathcal H}_{\text{int}}+{\mathcal H}_{C} &=& -\frac 15\left(
U+g\right) \sum_j\sum_{a>b}\left( {\mathcal M}_{j,ab}^{(1)2}+{\mathcal M}%
_{j,ab}^{(2)2}\right) \nonumber \\
&&-\frac U5\sum_j\sum_{a>b}2{\mathcal M}_{j,ab}^{(1)}{\mathcal M}_{j,ab}^{(2)},
\label{2H}
\end{eqnarray}
up to an additive constant. In the diagrammatic calculations, the vertices 
${\mathfrak T}^{\left( 1\right) }$ and ${\mathfrak T}^{\left( 2\right) }$,
corresponding to the first and the second terms in Eq.~(\ref{2H}) respectively, 
satisfy the identities ${\mathfrak T}^{\left( z\right) }={%
\mathfrak T}\circ {\mathcal U}^{\left( z\right) }$, $z=1,2$, where $\left( %
A\circ B\right) ^{\left( \gamma_1,\gamma _2;\gamma _3,\gamma _4\right) }%
\equiv \sum_{\beta
_1\beta _2}A^{\left( \gamma _1,\gamma _2;\beta _1,\beta _2\right) }B^{\left(
\beta _1,\beta _2;\gamma _3,\gamma _4\right) }$, and ${\mathcal U}^{\left(
z\right) \left( \gamma _1,\gamma _2;\gamma _3,\gamma _4\right) }$ are
antisymmetric with respect to the interchange $\gamma _1\leftrightarrow
\gamma _2$ and $\gamma _3\leftrightarrow \gamma _4$ and are defined by the
components: 
\begin{eqnarray*}
{\mathcal U}^{\left( 1\right) \left( 12;12\right) } &=&\frac 12,\ {\mathcal U}%
^{\left( 1\right) \left( 34;34\right) }=\frac 12, \\
{\mathcal U}^{\left( 2\right) \left( 13;13\right) } &=&\frac 12,\ {\mathcal U}%
^{\left( 2\right) \left( 14;14\right) }=\frac 12, \\
{\mathcal U}^{\left( 2\right) \left( 23;23\right) } &=&\frac 12,\ {\mathcal U}%
^{\left( 2\right) \left( 24;24\right) }=\frac 12.
\end{eqnarray*}
The rest of the components of ${\mathcal U}^{\left( z\right) }$ that remain 
undetermined after antisymmetrization are zero.
Then it follows that 
${\mathcal U}^{\left( z\right) }$ and ${\mathfrak T}^{\left( z\right) }$ %
have the following properties: ${\mathcal U}^{\left( 1\right) }\circ {%
\mathcal U}^{\left( 2\right) }=0$, ${\mathfrak T}^{\left( z\right) }\circ 
{\mathcal U}^{\left( z\right) }={\mathcal U}^{\left( z\right)
}\circ {\mathfrak T}^{\left( z\right) }={\mathfrak T}^{\left( z\right) }$, 
${\mathfrak T}^{\left( z\right) }\circ {\mathfrak T}^{\left( z\right) }
=\left( U^{(z)}/5\right) ^2{\mathcal U}^{\left( z\right) }$, where $%
U^{(1)}=U+g$ and $U^{(2)}=U$.

In order to calculate the critical temperature, we need to evaluate the
effect of the renormalization of ${\mathfrak T}_{ab}^{\left( z\right) }$ by the
next-order corrections. There are three of them in the second order, and for 
the purpose of the calculation of interest, the primary contribution is given 
by [Fig.\ \ref{Fig2}(b)]

\uprule

\begin{eqnarray}
\Gamma _{ab}^{\left( \gamma _1,\gamma _2;\gamma _3,\gamma _4\right) }\left(
k_1,k_2;k_3,k_4\right) &=&k_BT\sum_{\omega _q}\int \frac{d^2q}{\left( 2\pi
\right) ^2}\sum_{\beta _1\beta _2}{\mathfrak T}_{ab}^{\left( \beta _1,\beta
_2;\gamma _3,\gamma _4\right) }\left( q,s-q;k_3,k_4\right) \nonumber \\
&&\times {\mathfrak T}_{ab}^{\left( \gamma _1,\gamma _2;\beta _1,\beta _2\right)
}\left( k_1,k_2;s-q,q\right) {\mathfrak G}_{\beta _1\beta _1}^{\left( 0\right) 
}\left( q\right) {\mathfrak G}_{\beta _2 \beta _2}^{\left( 0\right) }
\left( s-q\right) ,
\end{eqnarray}

\downrule

\noindent 
where ${\mathbf s}={\mathbf k}_1+{\mathbf k}_2={\mathbf k}_3+{\mathbf %
k}_4$ and $\omega _s=\omega _1+\omega _2=\omega
_3+\omega _4$ are small. By considering simultaneously the corrections to
the vertex $\left( U/5\right) {\mathcal U}^{\left( z\right) }$ and taking
into account that ${\mathcal H}_{\text{int}}$ includes the sum over 15
generators [Eq. (\ref{expansion})], we find the expressions for the complete
vertices, ${\mathfrak T}_{\text{c }ab}^{\left( z\right) }={\mathfrak T}%
_{ab}^{\left( z\right) }/\left( 1-\sqrt{3/5}U^{(z)}\kappa ^{\left( z\right)
}\right) $, $z=1,2$, where
\begin{mathletters}
\begin{eqnarray}
\kappa ^{\left( 1\right) } &=&\frac 12\int \frac{d^2q}{\left( 2\pi \right) ^2%
}\frac{\tanh \left( \frac{\varepsilon \left( {\mathbf q}\right) -\mu _0}{2k_BT}%
\right) }{\varepsilon \left( {\mathbf q}\right) -\mu _0}, \label{eq_kappa1} \\
\kappa ^{\left( 2\right) } &=&\rho \left( \mu _0\right) .
\end{eqnarray}
\end{mathletters}

At critical temperature some of the complete vertices ${\mathfrak T}_{\text{c}%
}^{\left( z\right) }$ diverge. It takes place when either of the following
conditions is satisfied: 
\begin{mathletters}
\begin{eqnarray}
\left( U+g\right) \kappa ^{\left( 1\right) } &=& \sqrt{\frac 53}\text{, or} 
\label{eq_kappa21}\\
U\kappa ^{\left( 2\right) } &=& \sqrt{\frac 53}.  \label{eq_kappa2}
\end{eqnarray}
\end{mathletters}

The divergence of ${\mathfrak T}_{\text{c}}^{\left( 1\right) }$ can be regarded
as one of the Coulomb term ${\mathcal H}%
_{C}$, even if initially the latter was neglected. It implies that
the corresponding transition is antiferromagnetic. Therefore, 
Eq.~(\ref{eq_kappa21}) is the condition on AF transition and, similarly, 
Eq.~(\ref{eq_kappa2}) is one on dSC transition. The integral in the
right-hand side of Eq.~(\ref{eq_kappa1}) can be evaluated (see Appendix), 
which gives the value of the critical temperature of AF transition: 

\uprule

\begin{mathletters}
\begin{eqnarray}
k_BT_c &\simeq &2tD\exp \left( -\sqrt{\frac 35}\frac{\pi ^2}{\ln \left( 
\frac{2t}{\left| \mu _0\right| }\right) }\frac{2t}{U+g}\right) ,\ \frac{%
k_BT_c}{2t}\ll \frac{\left| \mu _0\right| }{2t}\ll 1, \\
&\simeq &2tD\left( \frac{\left| \mu _0\right| }{2t}\right) ^{3/8},\qquad 
\frac{\left| \mu _0\right| }{2t}\ll \frac{k_BT_c}{2t}\ll 1, \\
&\simeq &\left( \frac 35\right) ^{1/2}\frac{U+g}4,\qquad \frac{k_BT_c}{2t}%
\gg 1,  \label{high_T}
\end{eqnarray}
\end{mathletters}

\downrule

\noindent 
where $D =\gamma 2^{1/4}/\pi ^{1/2}\approx 0.387$ and 
$\gamma \approx 0.577$ is Euler's constant. 
We see that if $\left| \mu _0\right| \simeq 2tD^{8/5}$, the critical
temperature can be as high as the difference between the Fermi level and the
zero energy level $\varepsilon _{\mathbf k}=0$. For
temperatures higher than $2t$ the critical temperature attains the value
Eq.~(\ref{high_T}), although this assumes that $U/2t\gg 1$ and therefore can
not be regarded as a rigorous solution. However, the latter result allows
one to make a qualitative conclusion that in the strong coupling limit the
critical temperature is of order of $U$.

Eq.~(\ref{eq_kappa2}) appears to be an equation on the critical value
of chemical potential $\mu _{0c}$ so that the transition takes place at
small values of $\mu _0$. This has to do with the fact that in one-loop
order the density of states is diverging at the zero energy level and this
is why Eq.~(\ref{eq_kappa2}) can be solved for $\mu _{0c}$ for
arbitrary weak interaction $U$. In next orders the density of states will
likely to become finite and the peak will decrease with temperature, 
therefore, the equation (\ref{eq_kappa2}) will evolve into one on the 
critical temperature for the dSC transition.

Finally, by explicitly diagonalizing the mean-field Hamiltonian,\cite{Fradkin} %
one can find the spectrum of the excitations below critical temperature and 
the gap. For the AF transition, i.e., when one of the components of the 
superspin 
varies as $\left\langle {\mathcal N}_j\right\rangle =-\left( 5/8U\right)
N_0\cos \left( {\mathbf Q}\cdot {\mathbf r}\right) $, where ${\mathbf Q}=\left(
\pi ,\pi \right) $, $j=1$, $2$, or $3$, the Hamiltonian eigenvalues are $%
E_{\pm }\left( {\mathbf k}\right) =\mu \pm \left( \varepsilon _%
{\mathbf k}^2+N_0^2/4\right) ^{1/2}$ (double degenerate). Similarly, for
dSC transition ($j=4$ or $5$), there are four branches, $E_{\pm }^{(1)}%
\left( {\mathbf k}\right) =\left[ \left( \mu \pm \varepsilon _{\mathbf k}%
\right) ^2+N_0^2/4\right] ^{1/2}$ and $E_{\pm }^{(2)}\left( {\mathbf k}%
\right) =-E_{\pm }^{(1)}\left( {\mathbf k}\right) $. In the
purely $SU\left( 4\right) $ case\cite{note} $\mu =g=0$, the gap is equal $%
N_0=\left( 8\pi /e\right) t\exp \left( -5\pi t/U\right) $.

%%%%%%%%%%%%%%%%%%%%%%%%%%%%%%%%%%%%%%%%%%%%%%%%%%
\section{Specific heat and electrical conductivity}

The calculation in the previous section shows that in the mean-field 
approximation for a certain interval of the values of chemical potential 
the system is in dSC state at all temperatures. Certainly, if we included 
higher orders into our calculation, there would appear finite critical 
temperature. Although, strictly speaking, we have not shown existence of a 
transition to the normal phase as temperature increases, we can nonetheless 
consider the study of such a phase as a separate problem.

The most interesting case is when the critical temperature for the dSC 
transition is of order or higher than hopping $t$. Then the normal state 
becomes automatically the high-temperature regime for the gas of fermions, 
which dramatically changes the temperature dependence of kinetic and 
thermodynamic quantities.

The high-temperature limit of specific heat can be computed using the formula:
\cite{Abrikosov}
\begin{eqnarray}
c_V &=&\frac 1{2\left( k_BT\right) ^2}\int \frac{d^2k}{\left( 2\pi \right) ^2%
}\int d\omega \ \omega ^2\text{\,Im}\left[ G_R^{-1}\frac{\partial G_R}{%
\partial \omega }\right] _{\omega =0}  \nonumber \\
&=&\frac 1{2\left( k_BT\right) ^2}\int d\varepsilon \ \varepsilon ^2\rho \left(
\varepsilon \right) ,
\end{eqnarray}
where $G_R\left( \omega ,{\mathbf k}\right) $ is the retarded Green function.
Thus, $c_V\propto \left( t/T\right) ^2$for $k_BT\gg t$.

The DC conductivity can be derived from Kubo formula. In the large relaxation 
time approximation $\tau t\gg 1$, 
\begin{eqnarray}
\sigma _{xx} &=&\frac{e^2}{2\hbar ^2a}\int_{-\infty }^{+\infty }d\varepsilon \
K\left( \varepsilon -\mu \right) \rho _1\left( \varepsilon \right) , \\
K\left( \varepsilon -\mu \right)  &\simeq &2\tau \sum_\alpha \int_{-\infty
}^{+\infty }d\omega \ \left( -\frac{\partial f_F}{\partial \omega }\right) 
\nonumber \\
&&\qquad \times A\left( \omega -Y_\alpha \left( \varepsilon -\mu \right) 
\right) ,  \nonumber \\
\rho _1\left( \varepsilon \right)  &=&\int \frac{d^2k}{\left( 2\pi \right) ^2}%
\left( \frac{\partial \varepsilon _{{\mathbf k}}}{\partial k_x}\right)
^2\delta \left( \varepsilon -\varepsilon _{{\mathbf k}}\right) ,  \nonumber
\end{eqnarray}
where the spectral weight $A\left( \omega \right) 
=-\frac 1\pi \text{\,Im}
\left( \omega +i/2\tau \right) ^{-1}$, $a$ is a lattice constant 
for the sublattices, and $\tau $ is the relaxation time.

In order to analyze the Hall effect, it is necessary to modify the kinetic
term in the Hamiltonian by taking into account the presence of weak magnetic 
field: 
\begin{eqnarray}
{\mathcal H}_{\text{kin}} &=&\sum_{\alpha \sigma }\int \frac{d^2k}{\left( 2\pi
\right) ^2}\left( -te^{ik_x}c_{\alpha \sigma {\mathbf k}}^{\dagger}c_{\alpha \sigma 
{\mathbf ,k}_x,k_y-b_\alpha }\right.  \nonumber \\
&&-te^{-ik_x}c_{\alpha \sigma {\mathbf ,k}_x,k_y
-b_\alpha }^{\dagger}c_{\alpha \sigma {\mathbf k}} \nonumber \\
&&\qquad \left. -2t\cos k_yc_{\alpha \sigma {\mathbf k}}^{\dagger
}c_{\alpha \sigma {\mathbf k}}\right) ,
\end{eqnarray}
where $b_a=Y_\alpha b$, $b=\left| e\right| a^2B_z/\hbar c$, and we have 
assumed that the carriers are
negative-charged. The $xy$ component of the conductivity tensor is 
\begin{eqnarray}
\sigma _{xy} &=&-\frac{2\left| e\right| ^3aB_z}{3\hbar ^4c}\int_{-\infty
}^{+\infty }d\varepsilon \ 2\tau K\left( \varepsilon -\mu \right) \rho _2\left(
\varepsilon \right) , \\
\rho _2\left( \varepsilon \right)  &=&\int \frac{d^2k}{\left( 2\pi \right) ^2}%
\left( \frac{\partial \varepsilon _{{\mathbf k}}}{\partial k_x}\right) ^2%
\frac{\partial ^2\varepsilon }{\partial k_y^2}\delta \left( \varepsilon
-\varepsilon _{{\mathbf k}}\right) ,  \nonumber
\end{eqnarray}
and the Hall coefficient $R_H=\sigma _{xy}/\sigma _{xx}^2$. In the
high-temperature limit $k_BT\gg t$, the derivative of the Fermi function $%
\partial f_F/\partial \omega \simeq 1/4k_BT$ and consequently, in the same 
approximation as in the previous section, we find
\begin{mathletters}
\begin{eqnarray}
\sigma _{xx} &=&\frac{e^2}{\hbar ^2a}\frac{t^2\tau }{k_BT}, \\
\sigma _{xy} &=&-\frac{8\left| e\right| ^3aB_z}{3\hbar ^4c}\frac{t^3\tau ^2}{%
k_BT}, \\
R_H &=&-\frac{8a^3B_z}{3\left| e\right| c}\frac{k_BT}t.
\end{eqnarray}
\end{mathletters}
As we can see, the Hall coefficient increases linearly with temperature. On the
contrary, in the low-temperature limit, we recover the usual linear
dependence of specific heat and weak parabolic dependence of DC resistivity
on temperature as well as no dependence of the Hall coefficient on temperature.

%%%%%%%%%%%%%%%%%%%%%%%%%%%%%%%%%%%%%%%%%%%%%%%%%%
\section{Conclusion}

I have studied the model on a 2D bipartite lattice, which is equivalent to the 
two-leg ladder, that includes interaction of fermions from opposite 
sublattices. The corresponding term in the Hamiltonian is $SU\left(4\right) %
$-invariant. The symmetry breaking factors include chemical potential and 
Coulomb interaction. Fermions that belong to different sublattices have 
opposite ``hypercharge'', a symmetry-related quantum number. In the absence of 
leg-to-leg hopping the total hypercharge of the system conserves.

The physical meaning of the components of the order parameter is slightly
different from those given in Ref.\ \onlinecite{Zhang}, primarily in that 
the ground
state of the ``antiferromagnetic'' phase is actually a density wave with
varying rung N\'{e}el vector. In the strong $SU\left( 4\right) $ interaction 
limit, the degeneracy of the ground state is higher than one of the standard 
Hubbard model and the Hamiltonian resembles one of $t-J$ model, but also 
includes a pair hopping term with operators similar to those introduced in 
Ref.\ \onlinecite{Sachdev}.

The condition on the phase transition has been evaluated in one-loop order
in weak interaction limit. It leads to the equation on critical temperature
for the AF transition. However, for the dSC transition, it depends on the
density of states at the Fermi level, but not on temperature. In the latter
case, the dependence on temperature will probably appear in the next orders
of perturbation, but the resulting value of critical temperature will be
still abnormally high. Furthermore, if we formally consider the strong
coupling limit in the derived formulae, the condition on dSC transition will
be always satisfied, except when the Fermi 
surface lies in the area where the density of states is smaller than a 
threshold value of order of inverse coupling coefficient. Since the 
density of states will have a drop at the zero 
energy level, the AF transition will occur near to half-filling and dSC 
transition away from half-filling. When chemical potential $\mu _0=0$, 
no transition at finite temperatures takes place due to the Mermin~--\ Wagner 
theorem.

The fact that the critical temperature of the dSC transition can be of order
or higher than hopping implies that in the normal state the gas of fermions
will no longer resemble the usual zero-temperature Fermi liquid with 
discontinuity of the distribution function at the Fermi surface. The 
distribution function will almost linearly decrease 
with energy, which will result in abnormal temperature dependence of all 
experimentally measured characteristics. In particular, specific heat will be 
inverse proportional to the square of the temperature, and the DC electrical
resistivity and the Hall coefficient will be linearly proportional to the
temperature. Such dependence agrees with experimental data on high-$T_c$
cuprates.\cite{Ginsberg}

I would like to thank Prof.~J.~Preskill, S.-C.~Zhang, and P.~Weichman for
useful discussions and K.~Yang for comments. This work is supported in part 
by U.~S.\ Dept. of Energy under grant no. DE-FG03-92-ER\ 40701.

%%%%%%%%%%%%%%%%%%%%%%%%%%%%%%%%%%%%%%%%%%%%%%%%%%
\appendix
\section*{}

In this Appendix the integral that appears in the right-hand side of Eq.~(\ref
{eq_kappa1}) is evaluated: 
\begin{equation}
I=\int_0^\pi \frac{dk_x}{2\pi }\int_0^\pi \frac{dk_y}{2\pi }\frac{\tanh
\left( \frac{\varepsilon \left( {\mathbf k}\right) -\mu _0}{2k_BT_c}\right) }{%
\varepsilon \left( {\mathbf k}\right) -\mu _0}.  \label{integral}
\end{equation}
First, we make a substitution $k_{+}=\left( k_x+k_y\right) /2$, $%
k_{-}=\left( k_x-k_y\right) /2$ and expand the energy in terms of $k_{+}$
about the Fermi level $\varepsilon \left( {\mathbf k}\right) =\mu _0$: 
\begin{eqnarray}
&&\varepsilon \left( {\mathbf k}\right) -\mu _0\simeq -\sqrt{\left( 4t\cos
k_{-}\right) ^2-\mu _0} \nonumber \\
&&\qquad \ \times \left[ k_{+}-\arccos \left( \frac{\mu _0}{4t\cos k_{-}%
}\right) \right] .
\end{eqnarray}

In the limit of $\mu _0/2t\rightarrow 0$, the integral $I$ diverges
logarithmically. The internal integral can be taken by parts, which results
in a logarithmic part and a convergent integral. In the latter the limits
can be replaced by $\pm \infty $. Then there will be a region of integration
at $k_x=0$, $k_y=\pi /2$ that will not be covered and a symmetric region
that will be covered twice. However, the expression in the integral takes
the same value in both regions, therefore the result remains unchanged and
after the limit replacement no corrections will be necessary: 

\uprule

\begin{eqnarray}
I=\frac 1{\pi ^2}\int_0^{\frac 12\arccos \left( \frac{\mu _0}{2t}-1\right) }%
\frac{dk_{-}}{\sqrt{\left( 4t\cos k_{-}\right) ^2-\mu _0}}\ln \Biggl\{ \left[
\pi +\arccos \left( \frac{\mu _0}{4t\cos k_{-}}\right) -k_{-}\right] \Biggr. 
\nonumber \\
\times \Biggl. \left[ \arccos \left( \frac{\mu _0}{4t\cos k_{-}}\right)
-k_{-}\right] \left[ \left( 4t\cos k_{-}\right) ^2-\mu _0\right] \left(
\frac \gamma \pi \frac 1{k_BT}\right) ^2\Biggr\} .
\eqnum{A3}
\end{eqnarray}

\downrule

\noindent 
Here Euler's constant $\gamma \approx 0.577$ and $\mu _0$ is assumed to be
positive. Furthermore, Eq.~(\ref{integral}) does not depend on the sign of 
$\mu _0$, thus, we can replace $\mu _0$ by $\left| \mu _0\right| $. The
asymptotic expansion at $\left| \mu _0\right| /2t\rightarrow 0$ is 
\begin{eqnarray}
I &=&\frac 1{4\pi ^2t}\ln \left( \frac{2t}{\left| \mu _0\right| }\right) \ln
\left( \frac{\gamma 2^{1/4}}{\pi ^{1/2}}\frac{2t}{k_BT}\right) \nonumber \\
&&-\frac 3{32\pi ^2t}\ln \left( \frac{2t}{\left| \mu _0\right| }\right) ^2
+O\left[ \left( \frac{\left| \mu _0\right| }{2t}\right) ^0%
\right] .
\eqnum{A4}
\end{eqnarray}

%%%%%%%%%%%%%%%%%%%%%%%%%%%%%%%%%%%%%%%%%%%%%%%%%%

\end{multicols}

%%%%%%%%%%%%%%%%%%%%%%%%%%%%%%%%%%%%%%%%%%%%%%%%%%
\narrowtext
\begin{table}
\caption{Classification of the eigenstates of $\hat{\mathcal H}_{\text{int}}$.}
\label{table2}
\begin{tabular}{c|cccccc}
state & ${\mathcal N}^2$ & ${\mathcal L}^2$ & $Q$ & $S_z$ & $Y$ & $E_{%
\text{int}}/U$ \\ 
\hline
$\left| \Omega \right\rangle $ & 20 & 0 & 0 & 0 & 0 & 0 \\ 
$\left( N_1+iN_2\right) \left| \Omega \right\rangle $ & 4 & 16 & 0 & 2 & 0 & 
0 \\ 
$\left( N_1-iN_2\right) \left| \Omega \right\rangle $ & 4 & 16 & 0 & -2 & 0
& 0 \\ 
$N_3\left| \Omega \right\rangle $ & 4 & 16 & 0 & 0 & 0 & 0 \\ 
$\left( N_4+iN_5\right) \left| \Omega \right\rangle $ & 4 & 16 & 2 & 0 & 0 & 
0 \\ 
$\left( N_4-iN_5\right) \left| \Omega \right\rangle $ & 4 & 16 & -2 & 0 & 0
& 0 \\ 
$c_{\uparrow }\left| \Omega \right\rangle $ & 5 & 10 & 1 & -1 & -1 & 1 \\ 
$c_{\downarrow }\left| \Omega \right\rangle $ & 5 & 10 & 1 & 1 & -1 & 1 \\ 
$d_{\uparrow }\left| \Omega \right\rangle $ & 5 & 10 & 1 & -1 & 1 & 1 \\ 
$d_{\downarrow }\left| \Omega \right\rangle $ & 5 & 10 & 1 & 1 & 1 & 1 \\ 
$c_{\uparrow }^{\dagger }\left| \Omega \right\rangle $ & 5 & 10 & -1 & 1 & 1
& 1 \\ 
$c_{\downarrow }^{\dagger }\left| \Omega \right\rangle $ & 5 & 10 & -1 & -1
& 1 & 1 \\ 
$d_{\uparrow }^{\dagger }\left| \Omega \right\rangle $ & 5 & 10 & -1 & 1 & -1
& 1 \\ 
$d_{\downarrow }^{\dagger }\left| \Omega \right\rangle $ & 5 & 10 & -1 & -1
& -1 & 1 \\ 
$\Psi E\Psi \left| \Omega \right\rangle $ & 0 & 0 & 0 & 0 & -2 & 4 \\ 
$\Psi ^{\dagger }E\Psi ^{\dagger }\left| \Omega \right\rangle $ & 0 & 0 & 0
& 0 & 2 & 4 \\
\end{tabular}

\end{table}
%%%%%%%%%%%%%%%%%%%%%%%%%%%%%%%%%%%%%%%%%%%%%%%%%%

\end{document}